# PARTICIPATION ANTICIPATING IN ELECTIONS USING DATA MINING METHODS

Amin Babazadeh Sangar[1], Seyyed Reza Khaze[2] and Laya Ebrahimi[3]

[1]Faculty of Computer Science and Information System, Universiti Teknologi Malaysia, Skoda JB 81310, Malaysia,
`bsamin2@live.utm.my`

[2]Department of Computer Engineering, Science and Research Branch Islamic Azad University, West Azerbaijan, Iran,
`khaze.reza@gmail.com`

[3]Department of Computer Engineering, Science and Research Branch Islamic Azad University, West Azerbaijan, Iran,
`medyapamela@gmail.com`

*Abstract*

*Anticipating the political behavior of people will be considerable help for election candidates to assess the possibility of their success and to be acknowledged about the public motivations to select them. In this paper, we provide a general schematic of the architecture of participation anticipating system in presidential election by using KNN, Classification Tree and Naïve Bayes and tools orange based on crisp which had hopeful output. To test and assess the proposed model, we begin to use the case study by selecting 100 qualified persons who attend in 11th presidential election of Islamic republic of Iran and anticipate their participation in Kohkiloye & Boyerahmad. We indicate that KNN can perform anticipation and classification processes with high accuracy in compared with two other algorithms to anticipate participation.*

## 1. INTRODUCTION

Anticipating the political behavior of people in elections can determine the future prospect of each country domestic and foreign policies and characterize domestic and international relationships. This prospect will considerably help the candidates to anticipate their success and also provide an opportunity to know about the public demands and national will about their selected subjects. So, by emphasizing to these factors, they can achieve their goals. In addition, politics and functions of industry, economics, market and other sectors are selected based on experts' views that will be determined by people. So, economical and other sectors activists will also use these anticipations related to performed anticipations of their policies and programs for the future to provide better efficiency and output for themselves.

 In general, anticipating the public political behavior will indicate the society future prospect in which every one in every class and position can experience, program and make policy about the future of job, politics, economics, military and other sectors by being informed about authorities, parties and available candidates' viewpoints about future virtually. As most political theorists are in trouble in anticipating public political behavior and aren't able to correct anticipation about these kinds of elections by social and political analyzing methods and tools, data mining methods will be able to do so by collecting necessary data and documents from previous political and electoral behaviors of people which resulted in discovering potential rules of these data and reach to an accurate anticipation. So, using data mining and related algorithms will be considerable help to high and accurate anticipation from people political behavior. Firstly, we discuss about using data mining for political applications in different parts of the world in this article. Then, we provide a method to anticipate public participation in presidential election by introducing data mining and its practical methodologies. Then, we begin to test and assess the proposed model by using the case study of public viewpoint in Kohkiloye & Boyerahmad province in Iran. Finally, the obtained results of anticipation and classification are discussed and estimated by using data mining algorithms.

## 2. PREVIOUS WORK

In the first research, the researchers not only pointed out that the neural networks are increasingly used to solve non-linear controlling problems but also discussed about solving the problem of anticipating election result by using this model. Firstly, the researchers are going to use and make Two-layer Perceptron neural network to anticipate election result in India and then begin to teach (learn) network. The writers emphasized that during educational process it is created the minimum disorder and it can be used to anticipate and less-disorder procedure [1]. In the second study, they talked about the assumption of emphasizing on principle concentration of political behavior which acted logically in election. They also discussed about applying this assumption when they vote. The researchers develop a comparative method in their research approach which applies the differences of voters' decision making sequential processes. The writers point out to the fact that the goal of this research is to discover the potential relations and regulations in public political behavior of Spanish voters' data in election. They begin to use data mining process which is capable of finding the potential knowledge of given data. By using decision-making Trees and its learning by using j48 algorithm, it is found that the Spanish voters elected according to sequential reasoning [2].

In the third research, it is discussed about the importance of data mining to get the potential data to provide solution of solving a certain problem. The writers focus their research based on a basic model to recognize the relationship of persons who are qualified to vote and those participate in it. Using data mining techniques through linear regression, they find out about the importance of population and during election by using this model in USA election [3]. In the fourth research, it is discussed about the instability in voters' options due to the low interval in holding election. They classified the voters' approaches to two factors of emotional-oriented and logical-oriented selections. Then, it is provided an average (mean) approach by using the combination of these two factors which seems logical. This education is performed based on Perceptron neural network and determine whether volunteers are opportunists or benevolent. It helps to solve the instability problem [4].

## 3. DATA MINING AND KNOWLEDGE DISCOVERY

Nowadays, developing digital media resulted in data storage technologies growth in databases and brings widespread capacities and volumes of databases in all over the world [5]. By the increasing rate of volume, using traditional methods to obtain useful and proper patterns of data was practically difficult and expensive and sometimes didn't find the potential patterns so the International Journal on Cybernetics & Informatics ( IJCI) Vol.2, No.2, April2013traditional ones become ineffective. By explosive growth of stored data in databases, the need to provide new tools which automatically find patterns and knowledge from databases is felt.

In the late 1980s, the concept of data mining is developed and during this decade it was increasingly improved. Basically, data mining is defined as a concept in which we can obtain useful findings of relationships, patterns, tendencies and potential relations by using automatic patterns and analyzing large amount of data and data banks which have meaningful and potential patterns [6]. Data mining is the complicated process of recognizing patterns and accurate, new and potentially useful models as a large amount of data in a way that can be perceivable for human beings. Analyzing data, anticipating and assessing via patterns, classifying, categorizing and establishing association can be done by data mining.

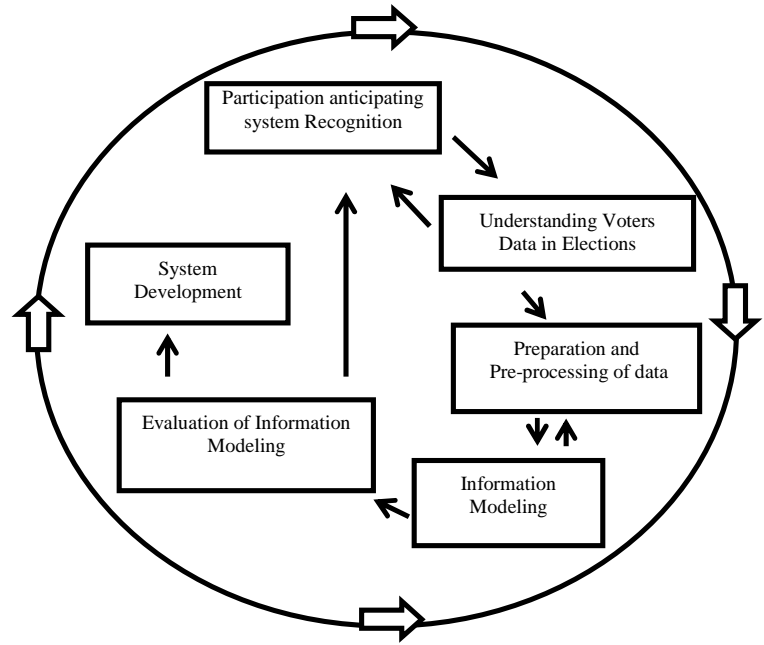

Figure 1.Project implementation plan based on CRISP data mining methodology

## 4. DATA MINING AND KNOWLEDGE DISCOVERY

There are many methods and methodologies to perform data mining projects which one of them is CRISP. CRISP is the abbreviation of CROSS INDUSTRY STANDARD PROCESS is consisted of system recognition, data preparation, and assessment and system development steps [7]. In Fig (1), it is indicated data mining project procedure based on this methodology.

Data mining algorithm is capable of anticipating, classification and clustering. In this section, we provide a general schematic of the architecture of anticipating system in presidential election by using Tools Orange, general architecture of typical data mining system and data mining project procedure based on CRISP which indicated in Fig (2).

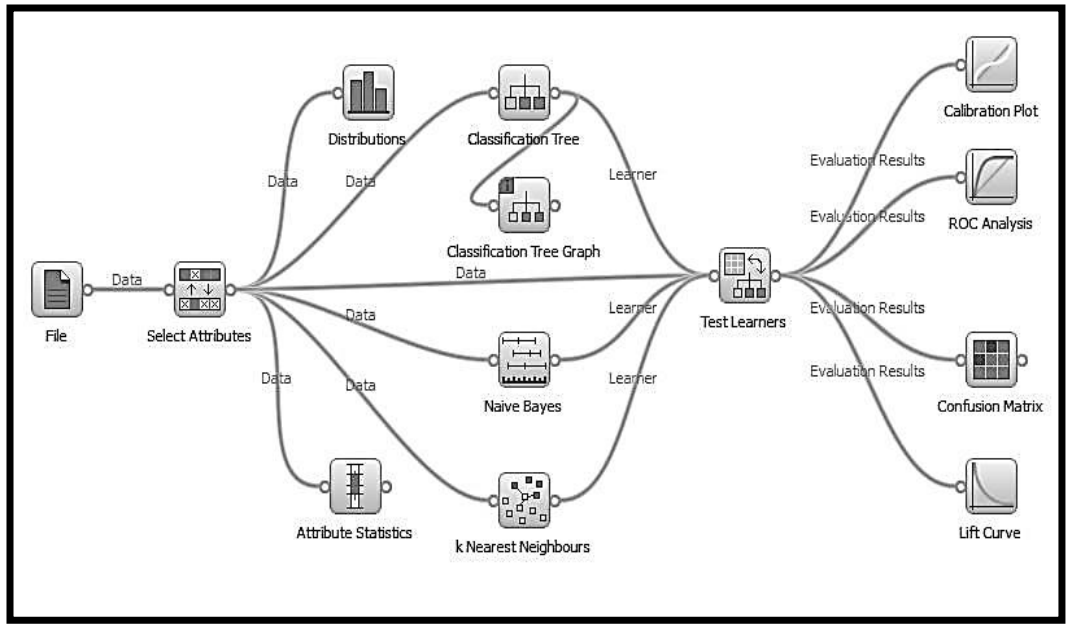

Figure 2.Forecasting system architecture using the Orange Tools

We use 3 basic data mining algorithms of KKN, Classification Tree and Naïve Bayes to classify and anticipate. In pattern assessment phase, we find out about the high accuracy of data mining pattern assessment standards. If it contains high anticipation and accuracy rate, we can use the performed anticipations to program.

## 5. PARTICIPATION ANTICIPATING USING THE PROPOSED MODEL: CASE STUDY

To test and assess the proposed model, we begin to use the case study by selecting qualified persons who attend in $11^{Th}$ presidential election of Islamic Republic of Iran and anticipate their participation in Kohkiloye & Boyerahmad. In the first phase, we begin to gather data. These data include characteristics and viewpoints of 100 qualified persons to attend election. The first characteristic is their age which classified to 3 groups of young, middle-aged and old. The second characteristic is their education which classified to four groups of PhD, MA, BA and Diploma. In this classification, the university students are also considered among these classes. The third characteristic is job and occupation which classified to 3 groups of government employee, clerks, university professors and tutors and self-employed occupations. In the fourth characteristics, we discuss about people political orientations (tendencies). In Iran, there are 3 main political orientations. Those who interested in basic public reforms in international relationships and economics are recognized as reformists and those who consider religious priorities and principles in international relationships and economics are called Fundamentalists. Of course, there is a third group, those who adopt Fundamentalists theory but are also obeying and follow the supreme religious and political leader of Iran and known as "Velayiees" or Moderate. We classified the people in these three groups. In the fifth characteristics, we are going to discuss about people view about government services consistency toward nation. In the presidency period of President Mahmoud Ahmadinejad, it was performed facilities to remove poverty in less-developed cities and districts which appreciated by people. The most important of these actions include housing mortgages, subsides reform which known as targeted subsides, marriage loans and fuel rationing.

50

We make questions about the importance of following activities in the future government and the main activities of it. In the sixth characteristics, we discuss about aspects and prospects to participate in Iran election. Some people consider it as a religious task and believe that participation in election is necessary to select Islamic government authorities according to the order of Islam and religious task. The others also believe that the goal of election is to participate in decision-making and help the democratic process. The other people think that the main goal is to perform general reforms to improve the current administrative procedure. It is clear that this characteristic has close relationship with people political prospect. The seventh characteristic involves the people prospect about country general politics in international affairs. In Iran and Middle-East political literature there is a term called resistance which means independency and deal with dependency of western developed country and their demands. Most people consider resistance as inseparable part of themselves and government political tasks due to their strong religious attitudes. Some others believe in international negotiations to solve Iran disputes and conflicts with the other countries. Iran international issues which create challenges are nuclear energy and how to enrich Uranium. Some people believe in resistance while the others prefer negotiation and compromise. We ask questions about this issue and gather their answers. In the eighth characteristics, it is asked about how to hold election and trust to the election officials. Some fully trust to the officials. However, the others believe in more supervision of Guardian Council due to the available conflicts and dispute of government with political groups and parties. Some others are somehow distrust to the election. In the ninth characteristics, we asked people views about the presidential volunteers. Some believe that they ought to be selected by people and nation. The others believe that they must be selected just by parties while the others think that political elites must attend in election due to the country particular circumstances. In the final characteristics, we asked them about participation in election process and gather their answers. It is the main characteristics of the people data bank and all the analyses are performed based on it. The obtained data are shown in Table (1).

Table 1.Data Table of the People Who can vote in elections

| No | Age | Degree | Job | Political Orientation | important task | Attitude to elections | Attitude to politics | Attitude to election officials | Attitude to candidates | Participation in elections |
|---|---|---|---|---|---|---|---|---|---|---|
| 1 | Old | Under license | free Job | Fundamentalists | Fuel Rationing | Religious duty | Negotiation | Unreliability | Party candidates | Without participation |
| 2 | Aged | Bachelor | Employee | Moderate | Marriage Loans | Religious duty | Resistance | Confidence | Popular candidate | Partnership |
| 3 | Old | Under license | free Job | Moderate | Mortgage | Religious duty | Resistance | Confidence | Popular candidate | Partnership |
| 4 | Aged | Bachelor | Employee | Fundamentalists | Targeted subsidies | Religious duty | Resistance | Higher accuracy | Popular candidate | Partnership |
| 5 | Old | Bachelor | Employee | Fundamentalists | Mortgage | Religious duty | Resistance | Confidence | Party candidates | Partnership |
| 6 | Old | Bachelor | free Job | Fundamentalists | Targeted subsidies | Religious duty | Resistance | Higher accuracy | Political elites | Partnership |
| 7 | Aged | Bachelor | free Job | Fundamentalists | Fuel Rationing | Religious duty | Resistance | Confidence | Party candidates | Partnership |
| 8 | Old | Bachelor | Employee | Reformist | Targeted subsidies | Reform | Negotiation | Confidence | Popular candidate | Without participation |
| 9 | Aged | Bachelor | free Job | Fundamentalists | Fuel Rationing | Religious duty | Resistance | Higher accuracy | Party candidates | Partnership |
| 10 | Aged | Bachelor | Employee | Fundamentalists | Marriage Loans | Religious duty | Resistance | Confidence | Popular candidate | Partnership |
| 11 | Aged | Under license | free Job | Reformist | Mortgage | Religious duty | Resistance | Unreliability | Popular candidate | Possible participation |
| 12 | Aged | Under license | Employee | Reformist | Targeted subsidies | Religious duty | Resistance | Confidence | Party candidates | Partnership |
| 13 | Aged | Under license | free Job | Fundamentalists | Fuel Rationing | Religious duty | Resistance | Confidence | Popular candidate | Partnership |
| 14 | Aged | Under license | free Job | Reformist | Mortgage | Partnership | Negotiation | Confidence | Popular candidate | Partnership |
| 15 | Aged | Under license | Employee | Reformist | Targeted subsidies | Religious duty | Resistance | Higher accuracy | Party candidates | Partnership |
| 16 | Young | Bachelor | free Job | Fundamentalists | Fuel Rationing | Religious duty | Resistance | Higher accuracy | Party candidates | Partnership |
| 17 | Aged | PhD | Employee | Fundamentalists | Targeted subsidies | Religious duty | Resistance | Unreliability | Popular candidate | Partnership |
| 18 | Young | Bachelor | free Job | Fundamentalists | Marriage Loans | Religious duty | Resistance | Confidence | Party candidates | Partnership |
| 19 | Young | Bachelor | free Job | Reformist | Fuel | Religious | Negotiation | Unreliability | Popular | Possible |

International Journal on Cybernetics & Informatics ( IJCI) Vol.2, No.2, April2013

| | | | | | Rationing | duty | | | candidate | participation |
|---|---|---|---|---|---|---|---|---|---|---|
| 20 | Aged | MA | Employee | Reformist | Targeted subsidies | Partnership | Compromise | Confidence | Popular candidate | Partnership |
| 21 | Young | Bachelor | free Job | Fundamentalists | Mortgage | Religious duty | Negotiation | Confidence | Party candidates | Partnership |
| 22 | Aged | PhD | Collegiate | Reformist | Fuel Rationing | Religious duty | Resistance | Confidence | Popular candidate | Partnership |
| 23 | Young | Bachelor | free Job | Fundamentalists | Mortgage | Partnership | Resistance | Confidence | Party candidates | Partnership |
| 24 | Aged | Bachelor | free Job | Fundamentalists | Targeted subsidies | Religious duty | Resistance | Confidence | Popular candidate | Possible participation |
| 25 | Aged | PhD | Collegiate | Moderate | Marriage Loans | Religious duty | Resistance | Confidence | Popular candidate | Partnership |
| 26 | Aged | Bachelor | free Job | Fundamentalists | Mortgage | Religious duty | Resistance | Higher accuracy | Party candidates | Partnership |
| 27 | Young | Bachelor | free Job | Fundamentalists | Mortgage | Religious duty | Resistance | Confidence | Party candidates | Partnership |
| 28 | Young | Bachelor | free Job | Reformist | Targeted subsidies | Reform | Compromise | Higher accuracy | Political elites | Partnership |
| 29 | Young | Bachelor | free Job | Fundamentalists | Mortgage | Religious duty | Resistance | Confidence | Popular candidate | Partnership |
| 30 | Young | Bachelor | free Job | Fundamentalists | Mortgage | Religious duty | Resistance | Confidence | Popular candidate | Partnership |
| 31 | Young | Bachelor | free Job | Fundamentalists | Targeted subsidies | Religious duty | Resistance | Confidence | Popular candidate | Partnership |
| 32 | Aged | MA | Employee | Reformist | Fuel Rationing | Partnership | Resistance | Unreliability | Party candidates | Possible participation |
| 33 | Old | Under license | free Job | Fundamentalists | Fuel Rationing | Partnership | Resistance | Higher accuracy | Party candidates | Partnership |
| 34 | Aged | Bachelor | Employee | Reformist | Targeted subsidies | Reform | Compromise | Confidence | Popular candidate | Without participation |
| 35 | Aged | Under license | free Job | Reformist | Mortgage | Reform | Resistance | Confidence | Popular candidate | Partnership |
| 36 | Aged | MA | Employee | Reformist | Marriage Loans | Reform | Negotiation | Higher accuracy | Political elites | Partnership |
| 37 | Young | PhD | Collegiate | Reformist | Mortgage | Reform | Resistance | Confidence | Party candidates | Partnership |
| 38 | Aged | Bachelor | Employee | Fundamentalists | Targeted subsidies | Religious duty | Negotiation | Confidence | Popular candidate | Partnership |
| 39 | Aged | Under license | free Job | Fundamentalists | Fuel Rationing | Religious duty | Resistance | Confidence | Popular candidate | Partnership |
| 40 | Old | PhD | Collegiate | Fundamentalists | Mortgage | Religious duty | Resistance | Higher accuracy | Party candidates | Partnership |
| 41 | Aged | Bachelor | Employee | Moderate | Mortgage | Partnership | Resistance | Confidence | Popular candidate | Partnership |
| 42 | Aged | Bachelor | free Job | Fundamentalists | Targeted subsidies | Religious duty | Resistance | Confidence | Party candidates | Partnership |
| 43 | Young | PhD | Collegiate | Reformist | Fuel Rationing | Reform | Compromise | Confidence | Popular candidate | Partnership |
| 44 | Aged | MA | Collegiate | Reformist | Mortgage | Reform | Resistance | Confidence | Popular candidate | Partnership |
| 45 | Aged | Under license | free Job | Reformist | Mortgage | Reform | Resistance | Confidence | Popular candidate | Partnership |
| 46 | Young | MA | Employee | Fundamentalists | Mortgage | Religious duty | Resistance | Confidence | Party candidates | Partnership |
| 47 | Young | MA | Collegiate | Fundamentalists | Targeted subsidies | Religious duty | Resistance | Confidence | Popular candidate | Possible participation |
| 48 | Aged | Under license | free Job | Reformist | Fuel Rationing | Reform | Resistance | Confidence | Party candidates | Partnership |
| 49 | Young | MA | Collegiate | Fundamentalists | Mortgage | Religious duty | Resistance | Confidence | Popular candidate | Partnership |
| 50 | Aged | MA | Employee | Reformist | Targeted subsidies | Reform | Resistance | Higher accuracy | Popular candidate | Partnership |
| 51 | Aged | Under license | free Job | Reformist | Marriage Loans | Reform | Resistance | Higher accuracy | Party candidates | Partnership |
| 52 | Young | Bachelor | Employee | Fundamentalists | Fuel Rationing | Religious duty | Negotiation | Unreliability | Party candidates | Partnership |
| 53 | Young | Bachelor | free Job | Moderate | Mortgage | Reform | Resistance | Higher accuracy | Popular candidate | Partnership |
| 54 | Young | Bachelor | free Job | Reformist | Targeted subsidies | Reform | Resistance | Unreliability | Popular candidate | Without participation |
| 55 | Aged | MA | Employee | Reformist | Mortgage | Reform | Negotiation | Confidence | Party candidates | Partnership |
| 56 | Young | Bachelor | Employee | Fundamentalists | Targeted subsidies | Religious duty | Resistance | Confidence | Popular candidate | Possible participation |
| 57 | Young | Bachelor | free Job | Fundamentalists | Marriage Loans | Religious duty | Resistance | Confidence | Party candidates | Partnership |
| 58 | Aged | Bachelor | Employee | Fundamentalists | Fuel Rationing | Religious duty | Resistance | Higher accuracy | Political elites | Partnership |
| 59 | Young | Bachelor | free Job | Fundamentalists | Targeted subsidies | Religious duty | Resistance | Confidence | Party candidates | Partnership |
| 60 | Young | Bachelor | Employee | Fundamentalists | Targeted subsidies | Religious duty | Resistance | Confidence | Popular candidate | Partnership |
| 61 | Young | Bachelor | free Job | Fundamentalists | Mortgage | Religious duty | Resistance | Confidence | Popular candidate | Partnership |
| 62 | Young | Bachelor | Employee | Reformist | Targeted subsidies | Reform | Resistance | Unreliability | Party candidates | Without participation |
| 63 | Young | Under license | free Job | Moderate | Fuel Rationing | Religious duty | Resistance | Higher accuracy | Popular candidate | Partnership |
| 64 | Aged | Bachelor | Employee | Fundamentalists | Mortgage | Religious duty | Resistance | Confidence | Party candidates | Partnership |
5252

| | | | | | | | | | | |
|---|---|---|---|---|---|---|---|---|---|---|
| 65 | Young | Under license | free Job | Fundamentalists | Targeted subsidies | Religious duty | Resistance | Confidence | Popular candidate | Partnership |
| 66 | Young | Under license | free Job | Reformist | Mortgage | Reform | Resistance | Confidence | Popular candidate | Partnership |
| 67 | Young | Under license | free Job | Reformist | Targeted subsidies | Religious duty | Resistance | Confidence | Popular candidate | Partnership |
| 68 | Young | Under license | free Job | Fundamentalists | Mortgage | Religious duty | Resistance | Confidence | Party candidates | Partnership |
| 69 | Young | Under license | free Job | Reformist | Marriage Loans | Reform | Resistance | Confidence | Political elites | Partnership |
| 70 | Young | Under license | free Job | Reformist | Targeted subsidies | Reform | Resistance | Higher accuracy | Party candidates | Partnership |
| 71 | Aged | MA | Collegiate | Reformist | Fuel Rationing | Religious duty | Compromise | Confidence | Party candidates | Partnership |
| 72 | Aged | Bachelor | Employee | Fundamentalists | Targeted subsidies | Reform | Resistance | Higher accuracy | Political elites | Partnership |
| 73 | Young | MA | Employee | Reformist | Marriage Loans | Reform | Resistance | Confidence | Party candidates | Partnership |
| 74 | Old | MA | Collegiate | Reformist | Mortgage | Partnership | Resistance | Confidence | Popular candidate | Without participation |
| 75 | Aged | Bachelor | Employee | Moderate | Mortgage | Religious duty | Resistance | Higher accuracy | Popular candidate | Partnership |
| 76 | Aged | PhD | Collegiate | Reformist | Targeted subsidies | Reform | Resistance | Confidence | Party candidates | Partnership |
| 77 | Young | MA | Collegiate | Fundamentalists | Mortgage | Partnership | Resistance | Confidence | Popular candidate | Partnership |
| 78 | Aged | Bachelor | Employee | Fundamentalists | Targeted subsidies | Religious duty | Resistance | Confidence | Popular candidate | Partnership |
| 79 | Young | MA | Collegiate | Reformist | Fuel Rationing | Partnership | Resistance | Higher accuracy | Party candidates | Possible participation |
| 80 | Young | MA | Employee | Fundamentalists | Marriage Loans | Religious duty | Resistance | Confidence | Popular candidate | Partnership |
| 81 | Old | MA | free Job | Fundamentalists | Mortgage | Religious duty | Resistance | Confidence | Popular candidate | Partnership |
| 82 | Young | MA | Collegiate | Reformist | Mortgage | Religious duty | Resistance | Confidence | Party candidates | Possible participation |
| 83 | Young | MA | free Job | Moderate | Targeted subsidies | Religious duty | Resistance | Confidence | Popular candidate | Partnership |
| 84 | Young | MA | Collegiate | Fundamentalists | Mortgage | Religious duty | Resistance | Confidence | Popular candidate | Partnership |
| 85 | Aged | Bachelor | Employee | Fundamentalists | Fuel Rationing | Religious duty | Resistance | Confidence | Party candidates | Partnership |
| 86 | Aged | MA | Employee | Moderate | Mortgage | Religious duty | Resistance | Confidence | Party candidates | Partnership |
| 87 | Young | MA | Collegiate | Fundamentalists | Targeted subsidies | Religious duty | Resistance | Confidence | Popular candidate | Partnership |
| 88 | Aged | Bachelor | free Job | Fundamentalists | Mortgage | Religious duty | Resistance | Confidence | Party candidates | Partnership |
| 89 | Young | Bachelor | Employee | Fundamentalists | Mortgage | Religious duty | Resistance | Confidence | Popular candidate | Partnership |
| 90 | Aged | Bachelor | free Job | Fundamentalists | Targeted subsidies | Religious duty | Resistance | Higher accuracy | Popular candidate | Partnership |
| 91 | Young | Bachelor | Employee | Reformist | Fuel Rationing | Reform | Negotiation | Unreliability | Party candidates | Possible participation |
| 92 | Aged | MA | Employee | Reformist | Targeted subsidies | Reform | Compromise | Confidence | Party candidates | Partnership |
| 93 | Young | Bachelor | Collegiate | Moderate | Marriage Loans | Religious duty | Resistance | Higher accuracy | Popular candidate | Partnership |
| 94 | Aged | Bachelor | Employee | Fundamentalists | Targeted subsidies | Reform | Resistance | Confidence | Party candidates | Partnership |
| 95 | Young | Bachelor | free Job | Fundamentalists | Mortgage | Religious duty | Resistance | Confidence | Political elites | Partnership |
| 96 | Aged | PhD | Collegiate | Reformist | Fuel Rationing | Religious duty | Resistance | Confidence | Popular candidate | Partnership |
| 97 | Young | Bachelor | Employee | Reformist | Targeted subsidies | Religious duty | Resistance | Higher accuracy | Party candidates | Partnership |
| 98 | Young | Bachelor | Employee | Fundamentalists | Mortgage | Partnership | Resistance | Confidence | Popular candidate | Partnership |
| 99 | Aged | Bachelor | free Job | Fundamentalists | Targeted subsidies | Religious duty | Resistance | Confidence | Popular candidate | Partnership |
| 100 | Young | Under license | Employee | Reformist | Fuel Rationing | Partnership | Negotiation | Higher accuracy | Party candidates | Possible participation |

After collecting data in the data visualization phase, we use distributed visualization which is provided in 3 to 11 Figures. It is necessary to note that all charts are discussed and reviewed due to the participation field in election. Participation in election is shown by blue color, possible participation by red and lack of participation by green, respectively.

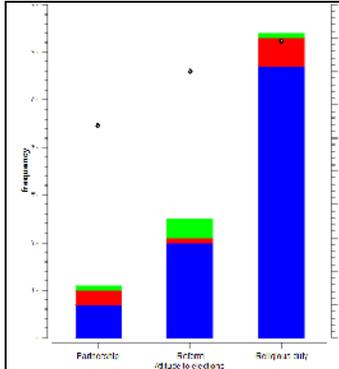

Figure 3. frequency of prospect to election and participation

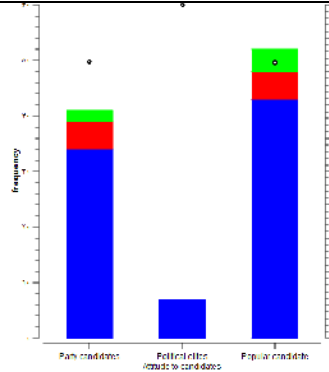

Figure 4. frequency of prospect to volunteers and participation

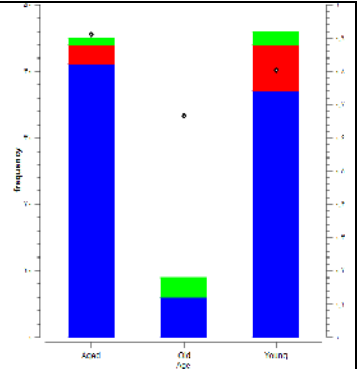

Figure 5. frequency of age and participation

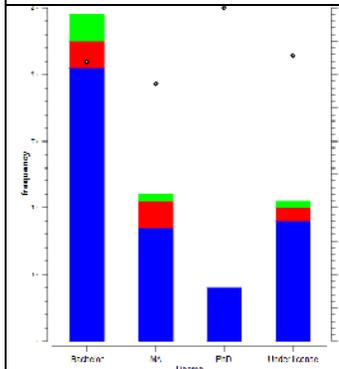

Figure 6. frequency of academic education and participation

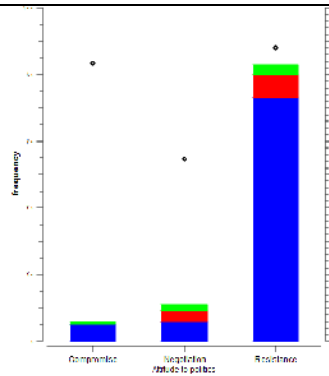

Figure 7. frequency of prospect to policy and participation

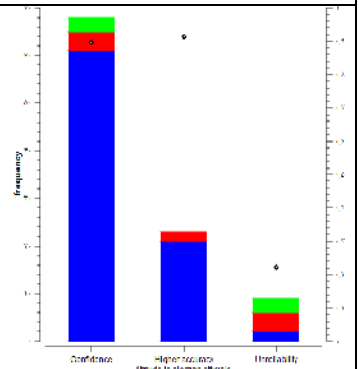

Figure 8. frequency of prospect to officials and participation

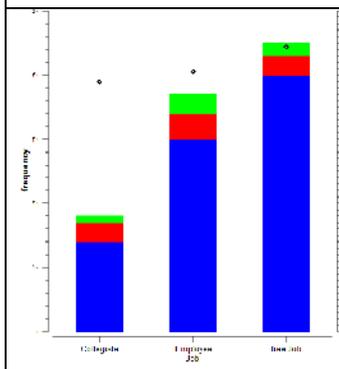

Figure 9. frequency of occupation and participation

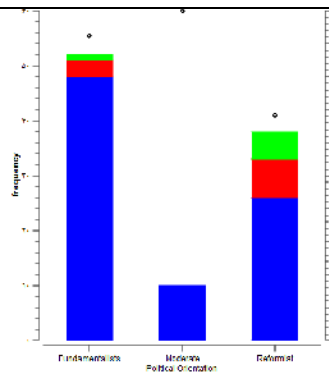

Figure 10. frequency of political attitudes and participation

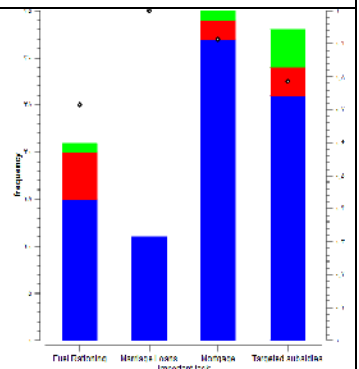

Figure 11. frequency of prospect to government tasks and participation

## 6. ANTICIPATION ALGORITHM TO PREDICT PARTICIPATION RATE IN ELECTION

In this section, we have a review on algorithms which are used in the proposed architecture.

### 6.1. Decision Trees

In artificial intelligence, it is used different states of concepts for better and clear representation by drawing decision making tree which make the audiences' learning, perceive and understanding easier and efficient. In fact, decision making tree is a type of proper tool and function to classify data, do estimation and provide anticipation due to the features and characteristics which data have till now [8]. We use unique methods of decision making trees which not only do classification but also make immediate decision-making easier and define the system properly. As it can be defined the system in the form of input and output set based on it, decision making tree can analyze features and outputs and provide the system as tree chart which resulted in data classification and system feature representation [9].

### 6.2. Naïve Bayes algorithm

One of the main tools to configure different techniques of data mining is Bayesian reasoning. Bayesian theory is a basic statistical method to solve problems of pattern recognition and classification. It establishes compromise among different classes' decisions and the costs of this decision and then chooses the best. So, for this kind of decision making, it is necessary to determine the possible distribution functions and their related values. Bayesian learning algorithms acts on different assumptions possibilities accurately. Naïve Bayes algorithm is a possible learning algorithm which is originated from Bayesian theory. It is a kind of classification which acts based on conditional possibilities of classes and involves two kinds: Bernoulli and Multi Nominal. The latter is proper as the size of database is big [10].

### 6.3. KNN algorithm

KNN method was firstly explained in 1950 and it was simple, efficient and applicable for few numbers of learning patterns and/or samples. So, it was used to recognize pattern. KNN is a method to classify objects based on the nearest educational samples in feature space which considered as the sample of instance-based or lazy learning. All the educational samples are saved firstly and as far as the unknown sample doesn't need classification, it won't be performed [11]. It is an efficient algorithm to classify and categorize.

## 7. DISCUSSION AND EVALUATION

In data mining, to review predictor accuracy, classification algorithm and anticipation, it is used concepts that we note them. Classifications accuracy is noted to the accuracy of classification and algorithms to anticipate new cases. Precision is the ratio of cases of a class of a case which classified accurately to all classified cases. Recall of a case is the ratio of cases of Class X which classified accurately to the number of class of that case. F-measure can be considered averagely as the weight of accuracy and integrity. Sensitivity is the number of positive labeled data which classified accurately to the all positive data and specificity is the number of negative labeled data which classified accurately to the all negative data [6, 12]. The complementary information is shown in 2-4 tables.

Table 2.The Anticipated Results of Participation by Using 3 Patterns of Data Mining Algorithm

| METHOD | CA | Sens | Spec | F1 | Prec | Recall |
|---|---|---|---|---|---|---|
| Classification Tree | 0.8300 | 0.9643 | 0.2500 | 0.9153 | 0.8710 | 0.9643 |
| KNN | 0.8700 | 0.9643 | 0.3750 | 0.9257 | 0.8901 | 0.9643 |
| Naive Bayes | 0.8600 | 0.9524 | 0.5000 | 0.9302 | 0.9091 | 0.9524 |

Table 3.The Anticipated Results of Possible Participation by Using 3 Patterns of Data Mining Algorithm

| NETHOD | CA | Sens | Spec | F1 | Prec | Recall |
|---|---|---|---|---|---|---|
| Classification Tree | 0.8300 | 0.2000 | 0.9667 | 0.2667 | 0.4000 | 0.2000 |
| KNN | 0.8700 | 0.4000 | 0.9667 | 0.4706 | 0.5714 | 0.4000 |
| Naive Bayes | 0.8600 | 0.4000 | 0.9667 | 0.4706 | 0.5714 | 0.4000 |

Table 4.The anticipated results of lack of participation by using 3 patterns of data mining algorithm

| METHOD | CA | Sens | Spec | F1 | Prec | Recall |
|---|---|---|---|---|---|---|
| Classification Tree | 0.8300 | 0.0000 | 0.9787 |  | 0.0000 | 0.0000 |
| KNN | 0.8700 | 0.3333 | 1.0000 | 0.5000 | 1.0000 | 0.3333 |
| Naive Bayes | 0.8600 | 0.3333 | 0.9681 | 0.3636 | 0.4000 | 0.3333 |

As it can be seen, in compared with two other algorithms, KKN algorithm can perform anticipation and classification with high accuracy. Naive Bayes also indicates better representation than Classification Tree. In reviews, it is used a chaos matrix in which the samples of a class, the other one or the similar one are shown. By using it, it can be seen certain samples which classified either accurate or wrong [14]. The matrixes related to these 3 algorithms are shown in 5-7 tables.

Table 5.chaos matrix of Naïve Bayes to anticipate participation in election

|  | Partnership | Possible participation | Without participation |  |
|---|---|---|---|---|
| Partnership | 80 | 2 | 2 | **84** |
| Possible participation | 5 | 4 | 1 | **10** |
| Without participation | 3 | 1 | 2 | **6** |
|  | **88** | **7** | **5** | **100** |

Table 6.chaos matrix of Classification Tree to anticipate participation in election

|  | Partnership | Possible participation | Without participation |  |
|---|---|---|---|---|
| Partnership | 81 | 1 | 2 | **84** |
| Possible participation | 8 | 2 | 0 | **10** |
| Without participation | 4 | 2 | 0 | **6** |

International Journal on Cybernetics & Informatics ( IJCI) Vol.2, No.2, April2013

| | | | | |
|---|---|---|---|---|
| | **93** | **5** | **2** | **100** |

Table 7.chaos matrix of KNN to anticipate participation in election

| | Partnership | Possible participation | Without participation | |
|---|---|---|---|---|
| **Partnership** | 81 | 3 | 0 | 84 |
| **Possible participation** | 6 | 4 | 0 | 10 |
| **Without participation** | 4 | 0 | 2 | 6 |
| | 91 | 7 | 2 | 100 |

The other index which is used to assess anticipations is Roc diagram. This model includes a diagram which indicates the relation among real positive and false positive rates. The false positive rate is negative topless which wrongly determined as positive and provided for a data model [15]. The results of Roc diagram are shown in 12-14 figures. In these diagrams and the next ones, KNN is shown in red, Naïve Bayes in blue and Classification Tree in green, respectively.

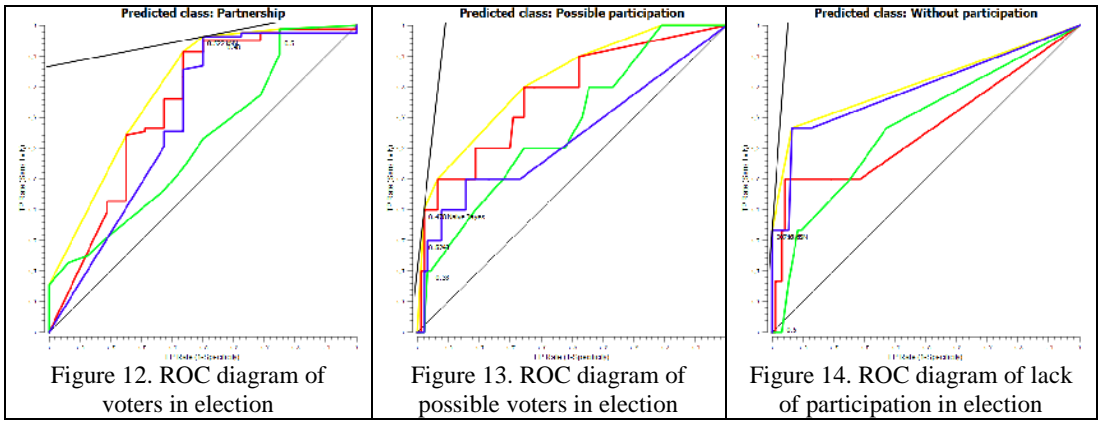

Figure 12. ROC diagram of voters in election    Figure 13. ROC diagram of possible voters in election    Figure 14. ROC diagram of lack of participation in election

The calibration plot also show the relationship between few cases which anticipated as positive and those were really positive [16]. It is shown 15-17 figures.

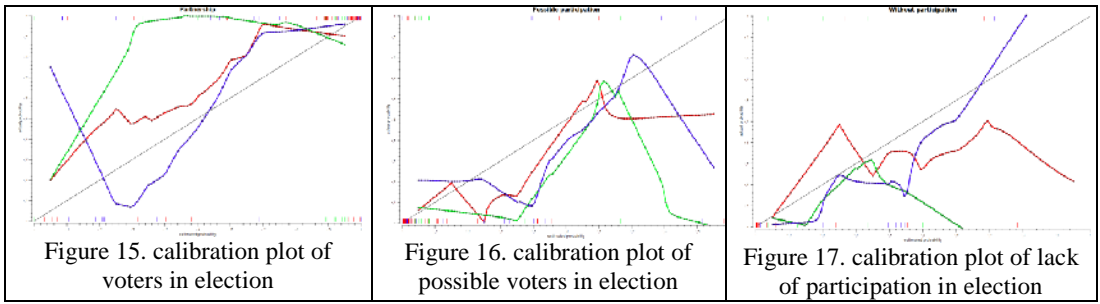

Figure 15. calibration plot of voters in election    Figure 16. calibration plot of possible voters in election    Figure 17. calibration plot of lack of participation in election

The other used index to assess is Lift curve which indicated the possibility of happening each event opposite of those which anticipated by classification [17]. In Fig 18-20, it is shown for these 3 algorithms.

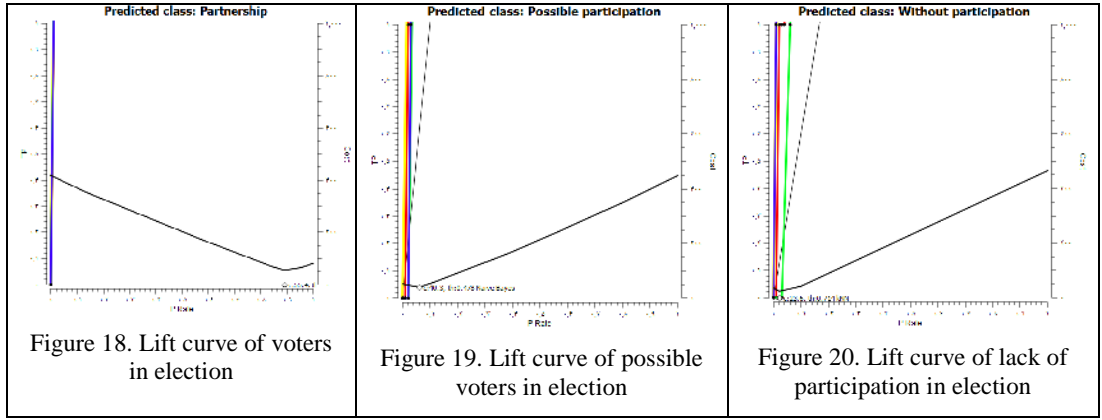

| Figure 18. Lift curve of voters in election | Figure 19. Lift curve of possible voters in election | Figure 20. Lift curve of lack of participation in election |

Decision making tree provides more convenient and proper representation for audiences than the other algorithms. The real and simple representation of anticipation is shown in Fig 21. The blue, green and red colors indicate those who attend election, non-attendants and possible participation, respectively. The main prospect is for the election official authorities. If there is full trust about those who hold election, the participation rate will be increased. If they demand more and detailed supervision to the election procedure or if they have consider it as a religious duty, then, they will attend it. But if they consider it as a participation to achieve democracy, their participation will be possible. However, if there is no trust to the authorities who hold the election, but the voter is a fundamentalist, he/she will certainly attend the election. Otherwise, if she/he is reformist and demands negotiation in international relationship, participation will be possible. At the other hand, if he/she demands compromise in international relationship and if he/she is young, he/she won't attend it otherwise the participation will be possible.

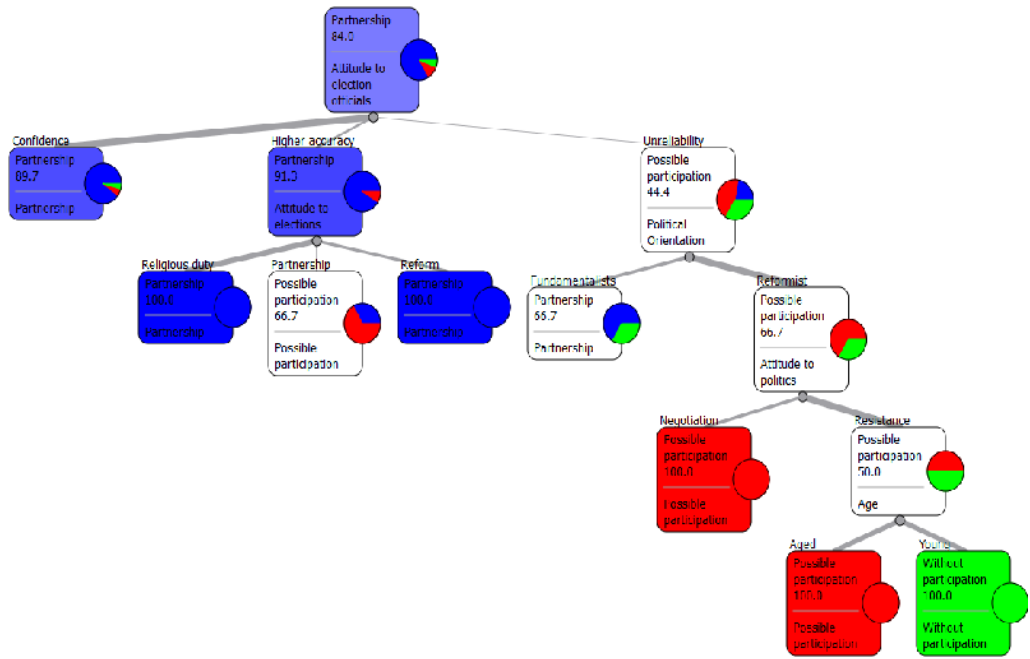

Figure 21. Classification Tree OF political behavior of people

## 8. CONCLUSIONS

Anticipating the political behavior of candidates can determine future of domestic and foreign policies of each country and their balances. Anticipating the political behavior of candidates will be considerable help for election candidates to assess the possibility of their success and to be acknowledged about the public motivations to select them. Nowadays, due to the increasing rate of data bases volume and inefficiency of traditional statistics methods to extract knowledge from data, it is used data mining algorithms to find potential relationships. It is capable of anticipating, classification and clustering. In this paper, we provide a general schematic of the architecture of participation anticipating system in presidential election by using Tools Orange based on CRISP and try to test and assess the proposed model. In this system, it is used 3 basic data mining algorithms of KKN, Decision-making Tree and Bayesian theory. . To test and assess the proposed model, we begin to use the case study by selecting 100 qualified persons who attend in $11^{Th}$ presidential election of Islamic Republic of Iran and anticipate their participation in Kohkiloye & Boyerahmad. We indicate that KKN in compared with Classification Tree and Naïve Bayes can perform anticipation and classification processes with high accuracy in compared with two other algorithms to anticipate participation.

**Authors**

**Amin Babazadeh Sangar** is doing his PhD at Information Systems Department, Faculty of Computer Science & Information Systems, Universiti Teknologi Malaysia,Skoda JB 81310, Malaysia.His working experience: 1.Deputy of Public relations and International affairs of Urmia University Of Technology, 2.Administrator of Web Developing Group Lab. Of Urmia University Of Technology, 3.Lecturer of Islamic Azad University of Iran.

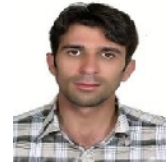

**Seyyed Reza Khaze** is a M.Sc. student in Computer Engineering Department, Science and Research Branch, Islamic Azad University, West Azerbaijan, Iran. His interested research areas are in the Operating Systems, Software Cost Estimation, Data Mining and Machine Learning techniques and Natural Language Processing. For more information please visit www.khaze.ir

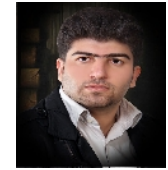

**Laya Ebrahimi** is a M.Sc. student in Computer Engineering Department, Science and Research Branch, Islamic Azad University, West Azerbaijan, Iran. His interested research areas are Meta Heuristic Algorithms, Data Mining and Machine learning Techniques.